\documentclass[preprint,12pt]{elsarticle}

\usepackage{epsfig}
\usepackage{amsthm}
\usepackage{amsmath,amsfonts,amssymb}
\usepackage{lineno}

\journal{Nuclear Physics B}

\begin{document}
\newcommand{\vek}[1]{\textbf{\textit{#1}}}
\newcommand{\as}{\underset{(\textbf{\textit{r}},t)}{\ast}}
\begin{frontmatter}

%% Title, authors and addresses

%% use the tnoteref command within \title for footnotes;
%% use the tnotetext command for theassociated footnote;
%% use the fnref command within \author or \address for footnotes;
%% use the fntext command for theassociated footnote;
%% use the corref command within \author for corresponding author footnotes;
%% use the cortext command for theassociated footnote;
%% use the ead command for the email address,
%% and the form \ead[url] for the home page:
%% \title{Title\tnoteref{label1}}
%% \tnotetext[label1]{}
%% \author{Name\corref{cor1}\fnref{label2}}
%% \ead{email address}
%% \ead[url]{home page}
%% \fntext[label2]{}
%% \cortext[cor1]{}
%% \address{Address\fnref{label3}}
%% \fntext[label3]{}

\bibliographystyle{elsarticle-num}
\title{The vector potential of a point  magnetic dipole}

%% use optional labels to link authors explicitly to addresses:
%% \author[label1,label2]{S. Sautbekov}
%% \address[label1]{}
%% \address[label2]{}

\author{S. Sautbekov}

\address{71 al-Farabi Ave., Almaty, Republic of Kazakhstan}

\begin{abstract}
The retarded vector potential of a point  magnetic dipole with an arbitrary time dependence undergoing accelerated relativistic motions is derived. A novel expression for the angular distribution of the radiated power of an arbitrary moving magnetic dipole is obtained. In particular, the case of an uniformly accelerated particle with a constant magnetic moment is considered.
The resulting equations are verified by showing that the fields reduce to less general forms found in the literature.
%% Text of abstract
\end{abstract}

\begin{keyword}
%% keywords here, in the form: keyword 
{Dipole \sep magnetic moment \sep retarded   potential\sep  radiation.}

%% PACS codes here, in the form: \PACS code \sep code

%% MSC codes here, in the form: \MSC code \sep code
%% or \MSC[2008] code \sep code (2000 is the default)

\end{keyword}

\end{frontmatter}

%% \linenumbers

%% main text
\section{Introduction}

Investigation of the radiation properties of a moving dipole is of undeniable theoretical and practical interest. 
For instance, the Thomas-Bargmann-Michel-Telegdi classical equation \citep{RanjbarPhRev} is generally used as an analytical tool in the interpretation of the extremely precise measurements of the magnetic moments of fundamental particles.  Besides,  hypothesized early-universe Big Bang conditions allow for neutrino radiation cooling and provide an energy loss-mechanism for subsequent neutrino condensation  \citep{Morley}. Transition radiation, and Bremsstrahlung radiation from fast neutrons play a critical role in the case of plasma as well. It is interesting that spin radiation manifests itself at high electron energies and can actually be measured in modern electron accelerators. Vector potential is useful in solving the pressing issues as registration of high-energy neutrino, which moving in a periodical magnetic field inside the magnetic undulator, as well as generation of  free-electron laser radiation.

 In spite of the fact that investigation of the electromagnetic field of an arbitrary moving dipole belongs primarily to the domain of classical electrodynamics, but it has not been sufficiently studied. 

In classical electrodynamics an electromagnetic field generated by an arbitrary moving point charge is known to be determined by the
Li\'{e}nard-Wiechert potentials \citep{jackson, becker1964, panofsky}. The textbook by Becker \cite{becker1964} contains concise derivations very much in the spirit of the original
literature. For historical perspectives see the books of Whittaker \citep{whittaker}. 

These potentials underlie the theory of relativistic radiation of charged particles. The Li\'{e}nard-Wiechert field appears as the retarded solution to Maxwell's equations with an moving point source. 

 Unfortunately, a general deduction of the exact electromagnetic field of an arbitrary moving dipole via the traditional potential approach is lacking from most undergraduate and graduate textbooks \citep{jackson, Landau, feynman, panofsky}. Li\'{e}nard-Wiechert technique has not been widely used in solving the radiation problems of arbitrarily moving point sources. Due to the difficulty of getting exact solutions, numerical methods and approximation theoretical methods have been developed. The literature on calculations of the fields of a moving point dipole is very extensive, including \citep{Monag, ward, Heras98PR, Glenn, Teitelboim, Burton, Bordovitsyn1995, Bialas}. 
 
 One of them is the so called multipole expansion method  \citep{Landau, feynman, PhysRevA}. The resulting expressions for the fields are usually called Jefimenko’s equations because they appeared for the first time in the textbook by Jefimenko \citep{jefimenko}. Heras \citep{HERAS98} discussed Jefimenko’s equations in material media to
obtain the electric and magnetic fields of a dipole in arbitrary motion and has derived Jefimenko’s equations from Maxwell’s equations using the retarded Green function of the wave equation.

 In some cases, the authors obtain the exact  fields for the electric dipole by assuming a harmonic time dependence for the sources \citep{jackson} or by using a very particular model for the charge and current distributions \citep{feynman, Sommerfeld}. In other cases, the authors are interested only in the radiation fields \cite{Landau, panofsky}.
As far as we know, a general deduction of the exact electric and magnetic fields of moving dipoles with arbitrary time dependence can be found in the works \citep{HERAS98, kort}.

Here we discuss derivation of the retarded vector potential of far reaching importance for analytical calculations of the electromagnetic field of an arbitrary moving magnetic dipole  in greater detail.

\section{Problem}

It is required to find the vector potential $\vek A(\vek r, t)$ of a point magnetic dipole possessing a magnetic moment $\vek M$ moving in an arbitrary direction with velocity $\vek v$.
The particle's motion is given parametrically
\begin{equation}\label{eq.0}
\textbf{\textit{r}}=\textbf{r}(\tau).
\end{equation}
It is assumed that the dipole motion is “a priori” known and that the path dimensions are macroscopically measurable in order that the classical field theory be valid. 

\section{Magnetic moment of a point particle}

The magnetic moment of a distributed current  in a certain volume is presented in a convolution form 
\begin{equation} \label{eq.1}
\textbf{\textit{M}}=\frac12\int_{\mathbb{R}^3}
\textbf{\textit{j}}(\textbf{\textit{r}}')\times(\textbf{\textit{r}}-\textbf{\textit{r}}')d^3\textbf{\textit{r}}'=
\frac12 \textbf{\textit{j}}\times*\textbf{\textit{r}}, 
\end{equation}
where the symbol $\times$ means the vector product,  $*$ is a convolution over space.
    
Note that the magnetic moment $\vek M$ in (\ref{eq.1}) doesn't depend on the location of the observation point ($\vek r$) and so the integral of the current density ($\vek j$) is equal to zero due to $\vek r$ can be taken out from under the integral sign.

It is to be noted that the current  \cite{seil} 
\begin{equation}\label{eq.2}
\textbf{\textit{j}}(\textbf{\textit{r}})=-(\textbf{\textit{M}}\times
\nabla)\delta(\textbf{\textit{r}})
\end{equation}
is the solution to the equation (\ref{eq.1}), 
where $\delta$ is Dirac delta function.
Indeed, Eq. (\ref{eq.1}) is identically  satisfied if $\vek j$ in (\ref{eq.2}) substitute into (\ref{eq.1})  and use the convolution property \cite{vladimir} 
 \begin{equation}
D(f* g)\equiv(D f)* g,
\end{equation}
where $D$ is any differential operator.

The eddy current expression (\ref{eq.2}) at $\vek r=0$ can be generalized for the current lines according to the particle motion equation 
 $\textbf{\textit{r}}=\textbf{\textrm{r}}(t)$ 
\begin{equation}\label{eq.4}
\textbf{\textit{j}}(\textbf{\textit{r}}, t)=-\big(\textbf{\textit{M}}(t)\times
\nabla\big)\delta(\textbf{\textit{r}}-\textbf{\textrm{r}}(t)).
\end{equation}

The above expression can be written as
\begin{equation}\label{eq:4}
\textbf{\textit{j}}(\vek r, t)=\nabla\times
\big(\vek M\,\delta(\vek r-\textbf{\textrm{r}}(t)\big)\big),
\end{equation}
according to the following handy identity and Eq. (\ref{eq.1})
\begin{equation*}
\begin{gathered}
-(\vek M\times
\nabla)\,\delta\equiv\nabla\times
(\vek M\,\delta)-\delta\;(\nabla\times\vek M),\\
\nabla\times\vek M=\frac12\vek j\times*(\nabla\times\vek r)=0.
\end{gathered}
\end{equation*}

It is worth emphasizing that useful to use the relativistic formula \citep{Alex}  for determine the magnetic moment  $\vek M$ of a particle moving with velocity $\vek v$  via the magnetic  $\vek M_0$ and electric $\vek P_0$ moments in the rest frame 
\begin{eqnarray}
\vek M(\tau)=\vek M_0+(1/\gamma-1)(\vek M_0 \vek v(\tau))\vek v(\tau)/v^2+\vek P_0\times\vek v,\\
\gamma=1/\sqrt{1-v(\tau)^2/c^2}.\nonumber
\end{eqnarray}

\section{The vector potential}
The retarded vector potential 
\begin{equation} \label{eq.7}
\textbf{\textit{A}}(\textbf{\textit{r}}, t)=-\mu\psi
\underset{(\textbf{\textit{r}},t)}{\ast}
\textbf{\textit{j}}
\end{equation}
 can be represented by the convolution of current density $\vek j$ with the Green function \citep{vladimir}
\begin{equation}\label{eq.5}
\psi(\textbf{\textit{r}}, t)=-\frac1{4\pi r}\delta(t-r/c).
\end{equation}
Here the symbol $ \raisebox{3pt}{$\underset{(\textbf{\textit{r}},t)}{\ast}$} $
 means a convolution over space and time, $c$ is the velocity of light.

Substituting expressions (\ref{eq:4}) and (\ref{eq.5}) into (\ref{eq.7}) we represent  the convolution in the integral form 
\begin{eqnarray} \label{eq:8}
 \vek{A}(\vek r,t)=\frac{\mu}{4\pi}\nabla\times\int_{-\infty}^{\infty}
 \frac{\delta\big(t-t'-|\vek r-\vek r'|/c\big)}{|\vek r-\vek r'|} \nonumber\\
 \int_{\mathbb{R}^3} \vek M(t')\,\delta\big(\vek r'-\textbf{\textrm{r}}(t')\big)d^3\vek r'\,dt'.
 \end{eqnarray}

The integral over all space in (\ref{eq:8}) is easily computed using the delta function:
\begin{eqnarray} \label{eq:9}
 \vek{A}(\vek r,t)=\frac{\mu}{4\pi}\nabla\times\int_{-\infty}^{\infty}\vek M(t')\,
 \frac{\delta\big(t-t'-|\vek r-\textbf{r}(t')|/c\big)}{|\vek r-\textbf{r}(t')|} \,dt'.
 \end{eqnarray}

The delta function in (\ref{eq:9}) can be simplified for time integration as
\begin{equation*}
\delta\big(F(\vek r,t,t')\big)=\frac{\delta(t'-\tau)}{|\partial F(\vek r,t,t')/\partial t'|}_{t'=\tau},
\end{equation*}
where
\begin{equation}\label{eq:10}
 F(\vek r,t,t')=t-t'-|\vek r-\textbf{r}(t')|/c,
\end{equation}
$\tau$ is a single root of the equation
\begin{eqnarray}\label{eq:11}
 F(\vek r,t,\tau)=0.
\end{eqnarray}
It is to be noted that, the relation (\ref{eq:11}) expresses in implicit form a functional dependence of the electromagnetic wave retardation time 
 $t-\tau=|\vek r-\textbf{\textrm{r}}(\tau)|/c$
 on the location of a particle and observation point.

 Thus, calculating the integral (\ref{eq:9}) and  introducing the notations
\begin{equation}\label{eq.11}
\vek R\equiv \vek r-\textbf{\textrm{r}}(\tau),\quad \boldsymbol{\beta}\equiv\vek v/c,\quad
\vek v=\partial\textbf{\textrm{r}}(\tau)/\partial \tau
\end{equation}
we obtain the vector potential
\begin{equation}\label{eq:12}
 \vek{A}(\vek{r},t)=\frac{\mu}{4\pi}\nabla \times \frac{\vek{M}(\tau)}{R-\vek R \boldsymbol{\beta}}\quad (v<c).
\end{equation}

In order to calculate the vector potential, we provide the expression (\ref{eq:12}) in the form
\begin{equation}\label{eq.13}
 \vek{A}(\vek{r},t)=\frac{\mu}{4\pi}\Big(-\vek M\times \nabla\frac{1}{R-\vek R \boldsymbol{\beta}}+ \frac{1}{R-\vek R \boldsymbol{\beta}}\nabla\times\vek M\Big),
\end{equation}
taking into account that
\begin{eqnarray*}
\nabla R=\vek n(1-\vek v\nabla\tau)=\frac{\vek n}{1-\vek n\boldsymbol{\beta}},\quad \vek n\equiv\frac{\vek R}{R},\\
\nabla(\vek R\boldsymbol{\beta})=\boldsymbol{\beta}+
\nabla\tau\frac{\partial(\vek R\boldsymbol{\beta})}{\partial\tau}=\boldsymbol{\beta}-\vek n\frac{(\vek R\dot{\boldsymbol{\beta}}/c)-\beta^2}{1-\vek n\boldsymbol{\beta}}.
\end{eqnarray*}

Note that the gradient has a dynamic part in the forgoing formulas.

The gradient is calculated similarly
\begin{equation}\label{eq:13}
\nabla\frac{1}{R-\vek R \boldsymbol{\beta}}=\frac{1}{R^2(1-\vek n\boldsymbol{\beta})^2}
\Bigg(\boldsymbol{\beta}-\vek n\frac{(\vek R\dot{\boldsymbol{\beta}}/c)+1-\beta^2}{1-\vek n\boldsymbol{\beta}}\Bigg)
\end{equation}
as well as the curl
\begin{equation}\label{eq:14}
\nabla\times\vek M\equiv\nabla\tau\times\frac{\partial}{\partial\tau}\vek M=-\frac{\vek n\times\dot{\vek M}}{c(1-\vek n\boldsymbol{\beta})}.
\end{equation}

Here, the overdots mean differentiation with respect to time ($\tau$) 
as well as taken into account that
 $$
\dot{\vek R}=-\vek v, \quad
\dot{R}=-\vek n\vek v.
$$
The expression
\begin{equation*}\label{eq14}
\nabla\tau=-\frac{\vek n}{c(1-\vek n\boldsymbol{\beta})}
\end{equation*}
useful for calculating the dynamic component of a curl (\ref{eq:14}) can be obtained by using the quotient rule and implicit differentiation of $F$ in (\ref{eq:10})
\begin{equation}\label{eq:15}
\frac{\partial\tau}{\partial x}=-\frac{\partial F/\partial x}{\partial F/\partial \tau}=-\frac{R_x}{cR(1-\vek n\boldsymbol{\beta})}, 
\end{equation}
where
$$
\frac{\partial F}{\partial x}=-\frac{R_x}{cR},\quad 
\frac{\partial F}{\partial \tau}=-1+\vek n\boldsymbol{\beta}.$$
After substituting (\ref{eq:13}) and (\ref{eq:14}) into (\ref{eq.13}) 
 we finally obtain the vector potential of a particle with a magnetic moment $\vek M$ 
 \begin{eqnarray}\label{eq:16}
\vek{A}(\vek{r},t)=-\frac{\mu}{4\pi}\,\bigg( 
 \frac{\vek M}{R^2(1-\vek n\boldsymbol{\beta})^2}\times\Big(\boldsymbol\beta
 -\vek n\frac{1-\beta^2} {1-\vek n\boldsymbol{\beta}}\Big)+\nonumber\\
 \frac{\vek n}{cR(1-\vek n\boldsymbol{\beta})^2}\times\Big(\dot{\vek M}+\vek M\frac{\vek n\dot{\boldsymbol{\beta}}}{1-\vek n\boldsymbol{\beta}}\Big)\bigg)_{\tau}. 
\end{eqnarray}

The first summand in (\ref{eq:16}) corresponds to the quasi-static eigen-field of a particle, 
the later to the radiation field  of a dipole moving with the acceleration  ($\dot{\boldsymbol{\beta}})$
as well as radiation due to alternating magnetic moment $(\dot{\vek M})$.

This formula constitutes a new result of classical electrodynamics.

\section{The particle radiation field}

Bearing in mind the implicit dependence of $\tau$ on the coordinates in $\vek A$ and taking into consideration the identical representation of the curl's dynamic part 
(\ref{eq:14}) 
we determine the electromagnetic field of a particle in the form
\begin{equation}\label{eq:17}
\vek H=\frac1{\mu}\Big(\nabla\tau\times
\frac{\partial}{\partial \tau}\vek A+\nabla\times\vek A\Big), 
\end{equation}
\begin{equation}\label{eq.18}
\vek E=-\frac{\partial\tau}{\partial t}\,\frac{\partial}{\partial \tau}\vek A.
\end{equation}

The foregoing general expressions (\ref{eq:17}), (\ref{eq.18}) for magnetic and electric fields  an arbitrary time dependent moving dipole have three contributions, namely, the static  zone contribution, proportional to $1/R^3$, the intermediate zone contribution, proportional to $1/R^2$ and the far zone of radiation contribution, proportional to $1/R$.

Substituting (\ref{eq:16}) into (\ref{eq:17}) and retaining the far zone contribution, proportional to $1/R$, it is easy to obtain 
 \begin{eqnarray}\label{eq:18}
\vek{H}^{Rad}(\vek{r},t)=\frac{\vek n}{4\pi c^2}\times\frac{\vek n}{R(1-\vek n\boldsymbol{\beta)}^3}\times\bigg( 
\ddot{\vek M}+\vek M
\frac{(\ddot{\boldsymbol{\beta}}\vek n)}{1-\vek n{\boldsymbol{\beta}}} +\nonumber\\
\frac{3(\dot{\boldsymbol{\beta}}\vek n)}{1-\vek n{\boldsymbol{\beta}}}\Big(\dot{\vek M}+\vek M
\frac{(\dot{\boldsymbol{\beta}}\vek n)}{1-\vek n{\boldsymbol{\beta}}} \Big)
 \bigg)_{\tau}. 
\end{eqnarray}

Taking into account of the derivatives
\begin{equation}\label{eq18}
 \frac{\partial\tau}{\partial t}=\frac{1}{1-\vek n\boldsymbol{\beta}},\quad \frac{\partial \vek n}{\partial \tau}=c\,\frac{\vek n(\vek n \boldsymbol\beta)-\boldsymbol\beta}{R}=\vek n\times(\vek n\times \boldsymbol\beta)\frac{c}{R},
\end{equation}
similarly, we obtain the electric field expression
\begin{eqnarray}\label{eq:19}
\vek{E}^{Rad}(\vek{r},t)=\frac{1}{4\pi\varepsilon}\frac{\vek n}{R c^2(1-\vek n\boldsymbol{\beta)}^3}\,\times
\bigg( 
\ddot{\vek M}+(\ddot{\boldsymbol{\beta}}\vek n)
\frac{\vek M}{1-\vek n{\boldsymbol{\beta}}} +\nonumber\\
\frac{3(\dot{\boldsymbol{\beta}}\vek n)}{1-\vek n{\boldsymbol{\beta}}}\Big(\dot{\vek M}+(\dot{\boldsymbol{\beta}}\vek n)
\frac{\vek M}{1-\vek n{\boldsymbol{\beta}}} \Big)
 \bigg)_{\tau}.
\end{eqnarray}
This formula was first obtained in \cite{Heras94} and then used for estimation the radiated energy by the dipole surface of a sonoluminescent bubble \cite{Esquivel}.

Note that the vectors  $\vek E^{Rad}$, $\vek H^{Rad}$ and $\vek n$ 
form a right-handed triple because of the expression
\begin{equation}\label{eq20}
\vek H^{Rad}(\vek{r},t)=(\mu/\varepsilon)^{-\frac12}\,\vek n\times\vek E^{Rad}(\vek{r},t),
\end{equation}
with respect to the magnetic and electric fields formulae (\ref{eq:18}), (\ref{eq:19}).

\subsection{The angular distribution of the  radiation power from a moving particle}

Now we're going to calculate the angular distribution of radiated power of a particle.

The radiation energy incident per unit area upon a surface $\vek nd\sigma$  at the observation point  during the receiving time interval $dt$ is equal to 
 $(\vek n\vek S)d\sigma\,dt$. 
 Here the Poynting vector is given by
  $\vek S=\vek E\times\vek H$.  

Dividing the radiation energy by the solid angle element $d\Omega=d\sigma/R^2$ and radiation time $d\tau$ we find the angular distribution of the radiation power
\begin{equation}\label{eq:20}
\frac{dP}{d\Omega}=(\vek n\vek S)R^2\frac{dt}{d\tau}=
(\vek n\vek S)R^2(1-\vek n\boldsymbol{\beta}).
\end{equation}
Here, according to the formula (\ref{eq.18}),  $dt=(1-\vek n\boldsymbol{\beta})d\tau$ is the registration time of the radiated power
where the change in the distance  $(\vek n\vek v) d\tau$ of a particle from the observation point during the time $\tau$ is taken into account, which leads to the change in the registration time  $(\vek n\boldsymbol{\beta})d\tau$.

It is worth emphasizing that the difference of duration the radiation time $d\tau$ and receiving time  $dt$ leads to a change the frequency of the wave reception by virtue of "Doppler factor" $(1-\vek n\boldsymbol{\beta})^{-1}$.

The angular distribution of the  radiation power from a moving particle can be written
\begin{equation}\label{eq:22}
\frac{dP}{d\Omega}=\frac{(1-\vek n\boldsymbol{\beta})}{\mu c}(R\vek E^{Rad})^2
\end{equation}
with help of the Poynting vector
\begin{equation}\label{eq:21}
\vek S=\frac{\vek n}{\mu c}(\vek E^{Rad})^2
\end{equation}
in agreement with the expressions (\ref{eq20}) and (\ref{eq:20}).

Finally, using the electric field in (\ref{eq:19}), the expression (\ref{eq:22}) for the angular power distribution is written in the general form
\begin{eqnarray}\label{eq:24}
\frac{dP}{d\Omega}=\frac{\mu}{(4\pi)^2 c^3}\frac{1}{(1-\vek n\boldsymbol{\beta)}^5}\,
\bigg( 
\vek n\times\ddot{\vek M}+(\ddot{\boldsymbol{\beta}}\vek n)
\frac{\vek n\times\vek M}{1-\vek n{\boldsymbol{\beta}}} +\nonumber\\
\frac{3\dot{\boldsymbol{\beta}}\vek n}{1-\vek n{\boldsymbol{\beta}}}\Big(\vek n\times\dot{\vek M}+(\dot{\boldsymbol{\beta}}\vek n)
\frac{\vek n\times\vek M}{1-\vek n{\boldsymbol{\beta}}} \Big)
 \bigg)_{\tau}^2.
\end{eqnarray}

\subsubsection{Radiation from a uniformly accelerated particle}

Let's consider a simple case of the expression (\ref{eq:24}) for a particle with a constant magnetic moment and  uniformly accelerated linear motion
($\dot{\vek M}=0$, $\ddot{\vek M}=0$,  $\ddot{\boldsymbol{\beta}}=0$)
\begin{equation}\label{eq:27}
\frac{dP(t)}{d\Omega}=\frac{9\mu}{(4\pi)^2 c^3}\left(\frac{(\vek n\times\vek M)^2(\dot{\boldsymbol{\beta}}\vek n)^4}{(1-\vek n\boldsymbol{\beta)}^9}\right)_{\tau}.
\end{equation}

Here it should be remembered that the instantaneous power $dP$ radiated into the solid angle $d\Omega$ is determined at the time $\tau$, where $d\Omega$ corresponds to its plane angle $\alpha=\angle (\vek n \vek v (\tau))$.

While the particle  travels  the distance  $R\beta$ during the time $R/c$, the observation point  should be seen from the angle $\theta$ at the time $t$ and the distance $R_t$ to the observation point.
 
Therefore, $\alpha$ in the vector products in (\ref{eq:27}) 
\begin{equation*}
|\vek n\times\vek M|=M\sin\alpha, \quad (\vek n\boldsymbol{\beta})=\beta\cos\alpha, \quad (\vek n\dot{\boldsymbol{\beta}})=\dot{\beta}\cos\alpha
\end{equation*}
must be defined by $\theta$. 

The angular dependence on $\alpha$ can be redefined by the expression
\begin{equation}\label{eq:28}
\alpha=\theta-\arcsin(\beta\sin\theta),
\end{equation}
which follows from a system of simple trigonometric equations
\begin{equation*}
\begin{cases}
 \cos\alpha=\beta+R_t/R\,\cos\theta,\\
\sin\alpha=R_t/R\,\sin\theta.
\end{cases}
\end{equation*}

Taking into account the expression (\ref{eq:28}), from (\ref{eq:27}) we obtain
\begin{equation}\label{eq:29}
\frac{dP(t)}{d\Omega}=\frac{9\mu\dot{\beta}^4M^2}{(4\pi)^2 c^3}F(\theta,\beta),
\end{equation}
where
\begin{equation}\label{eq:30}
 F(\theta,\beta)=\frac{\sin^2\alpha\,\cos^4\alpha}{(1-\beta\cos\alpha)^9} \quad (\vek M\perp\vek v).
\end{equation}

\begin{figure}[h]\label{fig1}
\begin{minipage}[h]{0.45\linewidth}
\center{\includegraphics[width=1.3\linewidth]{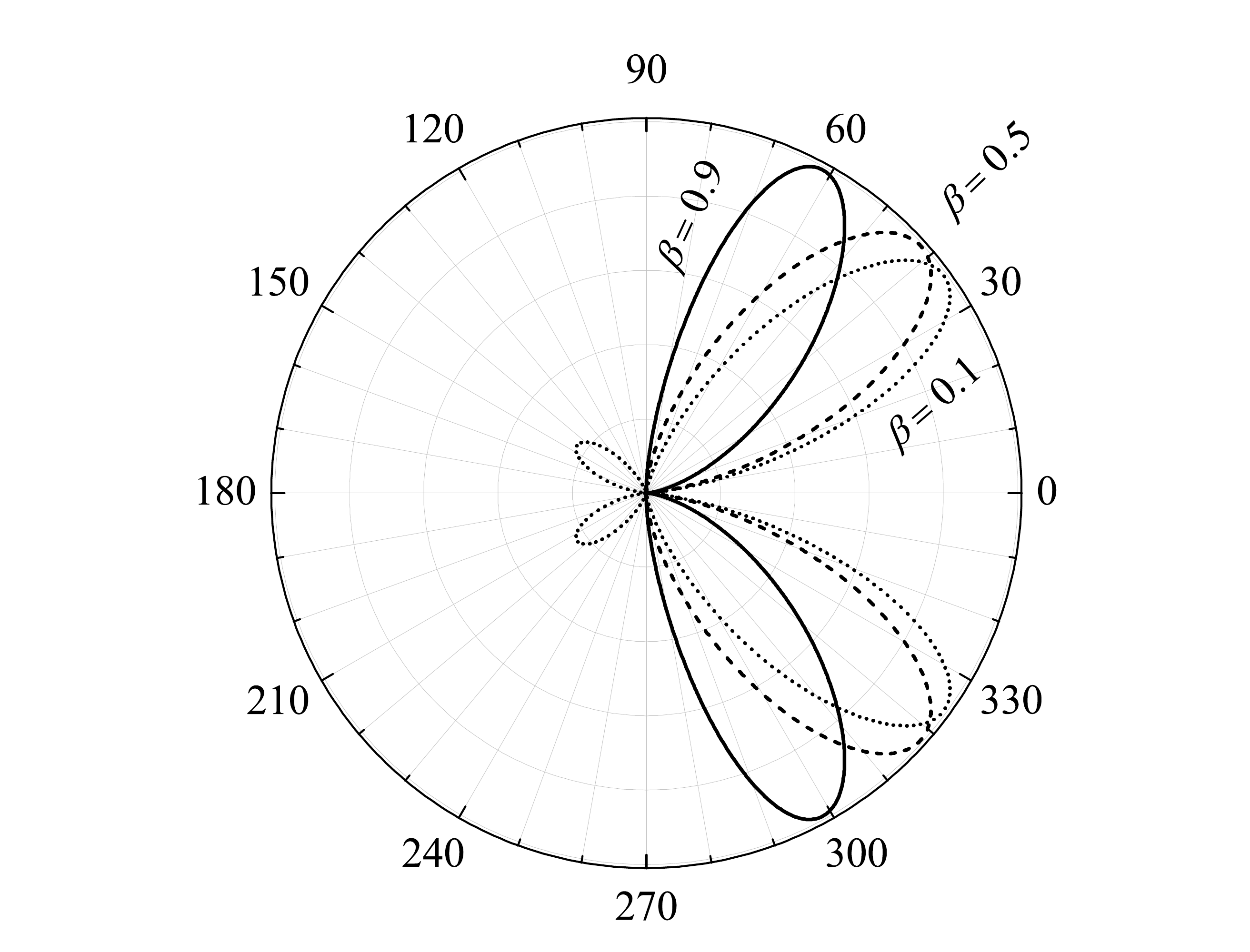} \\ a)   $\vek v\upuparrows\vek M$, $K(0.1)=0.3,\, K(0.5)=28.2,\, K(0.9)=0.9\cdot10^7.$}
\end{minipage}
\hfill
\begin{minipage}[h]{0.5\linewidth}
\center{\includegraphics[width=1.1\linewidth]{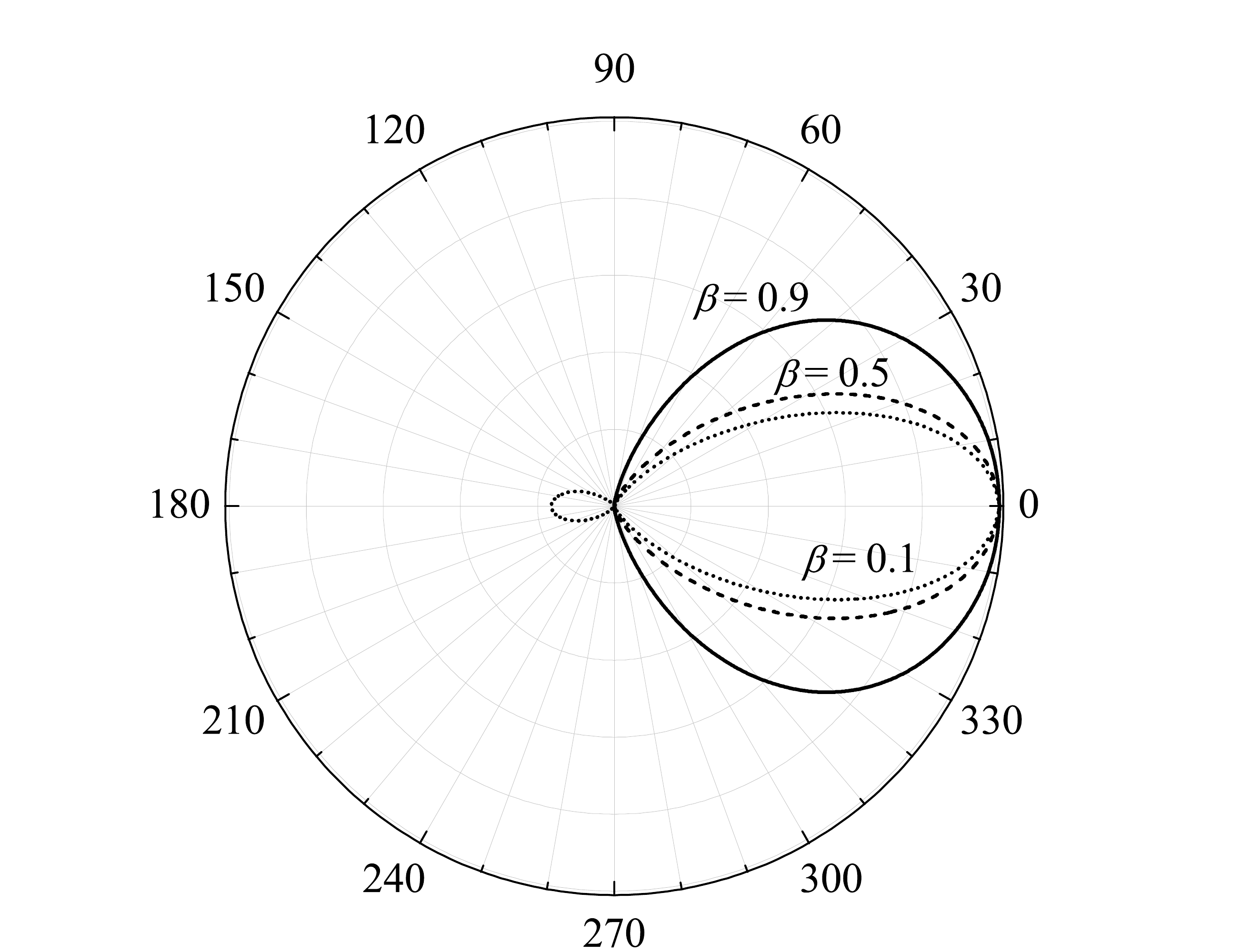} \\ b) $\vek v\bot\vek M$,\\ $K(0.1)=2.6,\, K(0.5)=512,\, K(0.9)=10^9$.}
\end{minipage}
\caption{The angular distribution $F(\theta,\beta)/K(\beta)$ of the radiation power of a magnetic dipole.}
\end{figure}

The normalized radiation pattern $F(\theta,\beta)/K(\beta)$ of a particle with the constant acceleration  $\vek a=\dot{\boldsymbol{\beta}}$ is presented in Figure (\ref{fig1}), where $K(\beta)=\max(F(\theta,\beta))$  is the maximum magnitude of radiation power. Here we consider  the following cases: $a$) the directions of the particle velocity and the magnetic moment are parallel ($\vek v\upuparrows\vek M$);

 $b$) the velocity and the  magnetic moment  are perpendicular  ($\vek v\bot\vek M$).
 
This is justified by the fact that the least radiation occurs towards the particle's motion  as well as by virtue of the inequality  $\theta >\alpha$ ($\theta \in[0,\pi]$) following from (\ref{eq:28}).
 
 In the case of a particle with a perpendicular magnetic moment, the radiation loss turns out by order magnitude greater (see Fig. \ref{fig1} $b)$), the maximum radiation is directed towards the motion of a particle.

The angular distribution of the  radiation power from a moving particle in this case is determined by the function
 \begin{equation}\label{eq:31}
 F(\theta,\beta)=\frac{\cos^6\alpha}{(1-\beta\cos\alpha)^9} \quad (\vek v\perp\vek M).
\end{equation}

\section{Discussion}

Let's look at same limiting cases. 
From the vector potential (\ref{eq:16}) in the limiting case $\beta \to 0$, it is easy to obtain the  vector potential of a nonrelativistic particle
 \begin{equation}\label{eq:31}
\vek{A}=\frac{\mu}{4\pi} 
 \frac{\vek M^{\star}\times\vek n}{R^2}, \quad  \vek M^{\star}=\vek M+\dot{\vek M}R/c.
\end{equation}
In particular, we have the well-known vector potential for a non-oscillating magnetic dipole \citep{feynman, hnizdo}
\begin{equation*}
\vek{A}=\frac{\mu}{4\pi} 
 \frac{\vek M\times\vek n}{R^2}.
\end{equation*} 

The exact expressions for electric and magnetic fields of a oscillating dipole at rest \citep{Heras98PR, kort} can be obtained from the formula (\ref{eq:31}) 
\begin{equation}\label{eq:32}
\begin{gathered}
4\pi\vek{H}=-\frac{4\pi}{\mu c}\vek n\times\dot{\vek A}+\nabla\times\frac{\vek M^{\star}\times\vek n}{R^2}= \\
\frac{3\vek n(\vek n\vek M^{\star})-\vek M^{\star}}{R^3}+\frac{(\ddot{\vek M}\times\vek n)\times\vek n}{c^2R},
 \end{gathered}
\end{equation}  
\begin{equation}\label{eq:33}
\vek E=-\dot{\vek A}=\frac{1}{4\pi\varepsilon}\frac{\vek n\times\dot{\vek M}^\star}{c^2R^2},
\end{equation}
it is sufficient to substitute (\ref{eq:31}) in (\ref{eq:17}) and (\ref{eq.18}).

\section{Conclusion}

 Explicit expression for the current density $\vek j$ (\ref{eq:4}) of an  arbitrary moving electrically neutral particle  possessing the magnetic moment $\vek M$ have been examined.
 
The general retarded vector potential $\vek A$  for a point dipole with an arbitrary time dependence magnetic moment $\vek M$ is obtained in (\ref{eq:16}) by way of  the current density $\vek j$ and Green function.
 
The formulae for the fields in  (\ref{eq:17}) and (\ref{eq.18}) is derived for an arbitrary moving point particle with a  magnetic moment oscillating in time, which expressed by the vector potential $\vek A$. 
As well as the general expression for the angular distribution of the radiated power of a magnetic dipole is obtained. 

The developed technique allows to calculate easily the electromagnetic field in the near and far zone. A direct deduction of the expressions for electric and magnetic fields (\ref{eq:19}), (\ref{eq20}) of a moving dipole in the far field zone is demonstrated. The first coincides with the exact expression obtained by J. Heras \citep{HERAS98} for the electric radiation field. The angular power distribution of the moving magnetic dipole is obtained, which coincides with the exact outcome of J. Heras work \citep{Heras98PR} possessing a constant magnetic moment. It is obvious that the above general expressions include the magnetostatic field \citep{Landau} in the limiting case $\beta\to 0$ as well as electromagnetic field of a oscillating dipole.
%% The Appendices part is started with the command \appendix;
%% appendix sections are then done as normal sections
%% \appendix

%% \section{}
%% \label{}

%% If you have bibdatabase file and want bibtex to generate the
%% bibitems, please use
%%
%\bibliographystyle{elsarticle-num} 
\bibliography{VecPot}
%% else use the following coding to input the bibitems directly in the
%% TeX file.

%%\begin{thebibliography}{VecPot}

%% \bibitem{label}
%% Text of bibliographic item

%\bibitem{}

%\end{thebibliography}
\end{document}